%Paper: hep-th/9407054
%From: kibler@lyolav.in2p3.fr
%Date: Mon, 11 Jul 94 16:31:50 +0200

  \magnification\magstep1
  \baselineskip = 0.5 true cm
  \parskip = 0.25 true cm

  \def\sa{\vskip 0.30 true cm}
  \def\degrees{$^\circ$}

  \pageno = 0
  \vsize = 22.5 true cm
  \hsize = 16.1 true cm

  \font\msym=msym10
  \def\gr{\hbox{\msym R}}

   \font\logo=logoipnl scaled \magstep4
   \line{\vbox{\hsize 3.5 true cm
   \noindent
   \logo P}\hfill \vbox{\hsize 3 true cm
   \noindent
   \null\hfill \bf LYCEN/9348\break
   \null\hfill October 1993}}

\sa
\sa
\sa
\sa
\sa
\sa
\sa

\noindent {\bf An $U_{qp}({\rm u}_2)$ model for rotational bands of nuclei}

\sa
\sa
\sa
\sa
\sa
\sa
\sa

\noindent {R Barbier, J Meyer and M Kibler}

\sa

\noindent {Institut de Physique Nucl\'eaire de Lyon,
IN2P3-CNRS et Universit\'e Claude Bernard,\break
43 Boulevard
du 11 Novembre 1918, F-69622 Villeurbanne Cedex, France}

\baselineskip = 0.68 true cm

\sa
\sa
\sa
\sa
\sa

\noindent {\bf Abstract}. A rotational model is developed
from a new version of the two-parameter quantum algebra
$U_{qp}({\rm u}_2)$. This model is applied to
the description of some recent experimental data for the
rotating superdeformed nuclei $^{192-194-196-198}{\rm Pb}$ and
                              $^{192-194        }{\rm Hg}$. A
comparison between the $U_{qp}({\rm  u}_2)$ model presented here
and the Raychev-Roussev-Smirnov             model with
                       $U_{q }({\rm su}_2)$ symmetry
shows the relevance of the
introduction of a second parameter of a ``quantum algebra'' type.

\sa
\sa
\sa
\sa
\sa
\sa
\sa
\sa
\sa
\sa
\sa
\sa

\noindent Submitted for publication in Journal
of Physics G: Nuclear and Particle Physics
\sa
\sa
\noindent Classification numbers: 2100, 2110, 2160E, 0220

\vfill\eject

\baselineskip = 0.91 true cm

\noindent
The concepts of quantum algebra (or Hopf algebra) and quantum group
(or compact matrix pseudo-group),
introduced in the eightees, continue to
be objects of considerable interest from
both a mathematical and a physical point of view. In nuclear physics,
quantum algebras have been applied
to rotational-vibrational spectroscopy of nuclei [1-6],
to the interacting boson model [7] and
to the ${\rm U}_3$ shell model [8].
In particular, a $q$-rotor model, based on the use of the quantum algebra
$U_q({\rm su}_2)$, has been developed by Raychev {\it et al}
[1] and, independently,
by Iwao [2]. This model has been successfully applied to the
description of rotational bands of deformed and superdeformed
nuclei. The connection between this $q$-rotor model and the VMI
(variable moment of inertia) model has been studied in great detail by the
Demokritos-Moscow-Sofia collaboration [3-5].

Most of the applications of quantum algebras to
(nuclear) physics have been restricted to the
use of one-parameter algebras, as for instance
$U_q({\rm su}_2)$, although several
theoretical works have been published in recent years
on multi-parameter, at least two-parameter,
quantum algebras. In this respect, we may quote among others the works
of references [9-15] mainly devoted to the
two-parameter algebra $U_{qp}({\rm su}_2)$.
However, it is well known (cf Drinfeld's theorem) that a
two-parameter deformation of a semi-simple Lie algebra turns out to be
essentially a one-parameter deformation. Therefore, in order to get
a nontrivial two-parameter deformation, that reduces to
$U_q({\rm su}_2)$ in some
limit, we have to deform the non semi-simple Lie algebra
${\rm u }_2$ instead of
${\rm su}_2$.

It is the aim of this letter to present a $qp$-rotor model
based on the two-parameter quantum algebra $U_{qp}({\rm u}_2)$
recently discussed in [16]. Such a model can be
applied to nuclear physics and to molecular physics as well. We
shall be concerned here with an application of the model to
rotational bands of superdeformed nuclei. In nuclear
physics, the motivation for constructing a $qp$-rotor model
having the $U_{qp}({\rm u}_2)$ symmetry is as follows.
The $q-$rotor model, based on the one-parameter algebra
$U_q({\rm su}_2)$, is especially
appropriate for describing rotational bands of deformed and superdeformed
nuclei at weak and medium angular momenta~; however, the latter model
is less convenient (as will be seen below) for describing those
 rotation energy levels,
at high angular momenta, which come from recent
experimental data in the $A \sim 190$ region. The objective of the present
work
is thus to develop a $qp$-rotor model and to test, on the
recent experimental results for the superdeformed (SD) bands of
    $^{192        -194}$Hg [17,~18]
and $^{192-194-196-198}$Pb [19-21],
the importance of introducing a second parameter when
passing from the $U_q   ({\rm su}_2)$ symmetry to the
                 $U_{qp}({\rm u }_2)$ symmetry.

The two-parameter deformation $U_{qp}({\rm u}_2)$ of the Lie
algebra ${\rm u}_2$ is spanned
by the four operators $J_\alpha$ ($\alpha = 0,3,+,-$) which satisfy the
commutation relations [16]
$$
[J_3,J_\pm   ] \; = \; \pm J_\pm                       \qquad
[J_+,J_-     ] \; = \; (qp)^{J_0-J_3} \; [[2J_3]]_{qp} \qquad
[J_0,J_\alpha] \; = \; 0.
\eqno {(1)}
$$
(In this letter, we use the notations
$[[X]]_{qp} = (q^X - p^X)/(q-p)$ and
$[X]_q \equiv [[X]]_{qq^{-1}} = (q^X - q^{-X})/(q-q^{-1})$,
where $X$ may stand for an operator or a number.) The co-product $\Delta_{qp}$
of the quantum algebra $U_{qp}({\rm u}_2)$ is defined by [16]
$$
\eqalign{
 \Delta_{qp}(J_\pm) \; = \;
&J_\pm \otimes (qp)^{{1\over 2}J_0} (qp^{-1})^{{1\over 2}J_3} +
 (qp)^{{1\over 2}J_0} (qp^{-1})^{-{1\over 2} J_3}
 \otimes J_\pm\cr
 \Delta_{qp}(J_3) \; = \;
&J_3 \otimes I + I \otimes J_3 \qquad \quad
 \Delta_{qp}(J_0) \;= \; J_0 \otimes I + I \otimes J_0 .\cr
}\eqno {(2)}
$$
The universal $R-$matrix, denoted here as $R_{qp}$ and
(partially) defined via $\Delta_{qp}(J_\alpha) =
                  R_{qp} \Delta_{pq}(J_\alpha)
                 (R_{qp})^{-1}$,
depends on the two parameters $q$ and $p$, as does
the co-product $\Delta_{qp}$.
For instance, the $R$-matrix corresponding to the coupling of two angular
momenta $j = {1\over2}$ reads (in terms of the unit matrices $E_{ab}$)
$$
R_{qp} \, = \, q \, (E_{11} + E_{44}) +
           {\sqrt {qp}} \, (E_{22} + E_{33}) +
           (q - p) \, E_{32}.
\eqno {(3)}
$$
It can be checked that the operator
$$
C_2 (U_{qp} ({\rm u}_2)) \; = \;
  {1 \over 2} \; (J_+ J_- + J_- J_+) +
  {1 \over 2} \; [[2]]_{qp} \; (qp)^{J_0-J_3} \; [[J_3]]^2_{qp}
\eqno {(4)}
$$
is an invariant of the quantum algebra $U_{qp}({\rm u}_2)$.
This invariant is the main mathematical ingredient for the
$qp$-rotor Hamiltonian model to be developed (see (10)).
For generic $q$ and $p$, each irreducible representation of
$U_{qp}({\rm u}_2)$
is characterized by a Young pattern
$[\varphi_1, \varphi_2]$ with $\varphi_1 - \varphi_2 = 2j$, where $2j$
is a nonnegative integer ($j$ will represent a spin angular momentum in what
follows)~; we note
$|[\varphi_1 , \varphi_2] m \rangle$
(with $m = - j, -j+1,\cdots,j)$
the basis vectors for the representation
$[\varphi_1, \varphi_2]$.
For physical reasons, we shall see in the following that $q = p^*$.
In this connection, it should be observed that the constraint
$q = p^*$ ensures that
$(\Delta_{qp} (J_\pm))^\dagger = \Delta_{pq}(J_\mp)$
and is compatible with the commutation relations for the operators
$\Delta_{qp} (J_\alpha)$ $(\alpha = 0,3,+,-)$.

Two particular cases are worth noticing. First, in the limiting situation
where $q = p^{-1} \rightarrow 1$, it is clear that equations
(1)-(4) reduce
to relations characterizing the Lie algebra ${\rm u}_2$. Second,
when $p = q^{-1}$
($q$ not being a root of unity), the three generators
$J_3$, $J_+$ and $J_-$ of the
algebra $U_{qq^{-1}}({\rm u}_2)$ generate the one-parameter quantum algebra
$U_{q}({\rm su}_2)$, ie the
usual deformation of the Lie algebra ${\rm su}_2$ introduced
in the pioneer works by Kulish, Reshetikhin, Sklyanin and
other authors (see references in [16]).
At this stage, we may wonder whether we
really gain something when passing from
 the ``classical'' quantum algebra
$U_q({\rm su}_2)$ to the quantum algebra $U_{qp}({\rm u}_2)$.
In this respect, let us define the operators $A_\alpha$
($\alpha = 0,3,+,-$) through
$$
J_\pm \, = \, (qp)^{{1\over 2}(J_0 - {1\over 2})} \, A_\pm
  \qquad \quad
  J_0 \, = \, A_0
  \qquad \quad
  J_3 \, = \, A_3
\eqno {(5)}
$$
and let us introduce
$$Q = (qp^{-1})^{1\over 2}
  \qquad \quad
  P = (qp)^{1\over 2}.
\eqno {(6)}
$$
Then, we can verify that the set
   $\{A_3, A_+, A_-\}$ spans $U_Q({\rm su}_2)$, which commutes
with $A_0$, so that we have the decomposition
$$
U_{qp}({\rm u}_2) = {\rm u}_1 \oplus U_Q({\rm su}_2).
\eqno {(7)}
$$
On the other hand, the invariant $C_2(U_{qp}({\rm u}_2))$ can be
developed as
$$
  C_2(U_{qp}({\rm u}_2)) \; = \;
  P^{2A_0-1} \; C_2(U_Q({\rm su}_2))
\eqno {(8)}
$$
where
$$
C_2(U_Q({\rm su}_2)) \; = \;
  {1 \over 2} \; (A_+A_- + A_-A_+) +
  {1 \over 2} \; [2]_Q \; [A_3]^2_Q
\eqno {(9)}
$$
is an invariant of $U_Q({\rm su}_2)$. Therefore,
in spite of the fact that the transformation (5)-(6)
allows us to generate the one-parameter algebra $U_Q({\rm su}_2)$ from the
two-parameter algebra $U_{qp}({\rm u}_2)$, see (7), the invariant
$C_2(U_{qp}({\rm u}_2))$,
as given by (8), still exhibits two independent parameters ($Q$ and $P$ instead
of $q$ and $p$).

We are now in a position to develop an $U_{qp}({\rm u}_2)$ model
for describing rotational bands of a deformed or superdeformed
nucleus. The first input of this model
lies on the use of the $qp$-rotor Hamiltonian
$$
H \; = \; { 1 \over 2{\cal I} } \; C_2 (U_{qp}({\rm u}_2)) + E_0
\eqno {(10)}
$$
where ${\cal I}$ denotes the moment of
inertia of the nucleus and $E_0$ the
bandhead energy. The second input
consists of choosing to diagonalise
$H$ on the subspace
$\{ | jm \rangle = | [2j,0] m \rangle \ : \ m = -j, -j +1, \cdots, +j \}$
of constant spin $j$ corresponding to the irreducible
representation for which $\varphi_1 = 2 j$ and $\varphi_2 = 0$.
Then, the eigenvalues of $H$ take the form
$$
  E \; = \; { 1 \over {2 {\cal I}} } \;
  [[j]]_{qp} \; [[j + 1 ]]_{qp} + E_0
\eqno {(11{\rm a})}
$$
or equivalently
$$
  E \; = \; { 1 \over {2{\cal I}} } \;
      {\rm e}^{ (2j-1) {{s+r} \over 2} } \;
      { {\sinh (j {{s-r}\over 2}) \;
         \sinh  [(j+1) {{s-r}\over 2}]} \over
       {\sinh^2 (  {{s-r}\over 2})} }
      + E_0
\eqno {(11{\rm b})}
$$
where $s = \ln q$ and
      $r = \ln p$.
Two important constraints can be imposed
on the parameters $s$ and $r$. Indeed,
{\it (i)} in the particular case
$p = q^{-1}$ (ie, $r = -s$),
we want that our model reduces to the
$U_q({\rm su}_2)$ model developed by Raychev {\it et al} [1] and
{\it (ii)} for obvious reasons, $E$ should be real. Points
{\it (i)} and
{\it (ii)} lead to the constraints
$ (s-r) \in {\rm i} \gr $ and
$ (s+r) \in         \gr $. Therefore, we shall assume
$$
  \matrix{
  {{s+r}\over 2        } = \beta \cos \gamma\cr
\cr
  {{s-r}\over 2 {\rm i}} = \beta \sin \gamma\cr
  }
  \quad
  \matrix{
  \Longleftrightarrow \cr
  }
  \quad
  \matrix{
  q = {\rm e}^{\beta \cos \gamma} \; {\rm e}^{+{\rm i} \beta \sin \gamma}\cr
\cr
  p = {\rm e}^{\beta \cos \gamma} \; {\rm e}^{-{\rm i} \beta \sin \gamma}\cr
  }
\eqno {(12)}
$$
where $\beta$ and $\gamma$ are two independent real parameters.
(Note that the parameters $q$ and $p$ defined by (12)
satisfy $q = p^*$.) By introducing (12) into (11b), we
obtain the rotational energy
$$
E \; = \; { 1 \over {2 {\cal I}} } \;
      {\rm e}^{(2j-1) \beta \cos \gamma} \;
      { {\sin (j \beta \sin \gamma) \; \sin [(j+1) \beta \sin \gamma]} \over
        {\sin^2 (\beta \sin \gamma)} }
      + E_0.
\eqno {(13)}
$$
This expression for  $E$  constitutes our basic
result for applications. In the particular case
where $\gamma = {\pi \over 2}$ (ie, $q = p^{-1} = {\rm e}^{{\rm i}\beta})$,
(13) coincides with the corresponding expression derived in [1].

Before testing formula (13) on
some recent experimental data,
we would like to show that our
$U_{qp}({\rm u}_2)$
model is not phenomenologically equivalent
(in the sense discussed in [3-5])
to the VMI model. As a matter of fact, if we try to obtain an
{\it ` la} Dunham expansion of $E$, we find
$$
E \; = \; { 1 \over {2{\cal I}_{\beta \gamma}} } \;
      \left(
\sum_{n=0}^\infty \;
      d_n (\beta,\gamma) \; [C_2 ({\rm su}_2)]^n +
      [2 C_1 ({\rm u}_1) + 1] \;
\sum_{n=0}^\infty \;
      c_n (\beta,\gamma) \; [C_2 ({\rm su}_2)]^n \right) +
      E_0
\eqno (14)
$$
where
$$
{\cal I}_{\beta \gamma} =
  {\cal I} \, {\rm e}^{2 \beta \cos \gamma} \quad \qquad
  C_2 ({\rm su}_2) =
  j(j+1) \quad \qquad
  C_1 ({\rm u}_1) = j
\eqno (15{\rm a})
$$
while the expansion coefficients $c_n(\beta, \gamma)$
                             and $d_n(\beta, \gamma)$
are given by the series
$$
\eqalign{
  c_n(\beta, & \gamma) \; = \;
  { 2^{2n} \over {2 \sin^2(\beta \sin \gamma)}} \cr
& \sum_{k=0}^\infty \;
  \{ (\cos \gamma)^{2k+1+2n}
  \cos (\beta \sin \gamma) - \cos [(2k + 1 + 2n)\gamma]\} \;
  {{\beta^{2k+1+2n}} \over {(2k+1+2n)!}} \;
  {{(k+n)!} \over {k! \, n!}}\cr
\cr
  d_n(\beta, & \gamma) \; = \;
  { 2^{2n} \over {2 \sin^2(\beta \sin \gamma)}}\cr
& \sum_{k=0}^\infty \;
  \{ (\cos \gamma)^{2k+2n}
  \cos (\beta \sin \gamma) - \cos [(2k + 2n)\gamma]\} \;
  {{\beta^{2k+2n}} \over {(2k+2n)!}} \;
  {{(k+n)!} \over {k! \, n!}}.
}
\eqno {(15{\rm b})}
$$
Note that for $\gamma = {\pi \over 2}$, we have
$c_n (\beta, {\pi \over 2}) = 0$ and the expression
(14) simplifies to
$$
E \; = \; {1 \over 2 {\cal I}} \;
      {{\beta^2} \over {\sin^2 \beta}} \;
      \sum_{n=1}^\infty \; (-1)^{n-1} \;
      {{2^{n-1}} \over {n!}} \;
      \beta^{n-1} \; j_{n-1}(\beta) \; [j(j+1)]^n + E_0
\eqno {(16)}
$$
in terms of the spherical
Bessel functions of the first kind $j_n$. The particular
expansion (16) is in agreement with the result of Bonatsos
{\it et al} [3] (see also [4,~5]). The
more general expansion (14) provides us with a development of $E$ in terms of
the second-order invariant $C_2({\rm su}_2)$ of the Lie algebra
${\rm su}_2$,
which invariant characterizes the energy of a rigid rotor,
and of the first-order invariant
$C_1({\rm u}_1)$
of the Lie algebra ${\rm u}_1$. The expansion (14)
differs from the ones
for the VMI model and for the $U_q({\rm su}_2)$ model
by the occurrence of
$C_1({\rm u}_1)$.
We thus expect that our $U_{qp}({\rm u}_2)$ model leads to results
different from the ones obtained by using the VMI model or the
$U_q({\rm su}_2)$ model.

We have applied the energy formula (13) to the
SD bands which appear for some
nuclei in the $A \sim 190$ region. We report here on
the analysis of seven SD bands in six even-even
nuclei, namely $^{192-194        }$Hg [17,~18] and
               $^{192-194-196-198}$Pb [19-21]. In the framework of
our $U_{qp}({\rm u}_2)$ model, the two parameters
$\beta \sin \gamma$ and
$\beta \cos \gamma$ occurring in
(13) are taken as free independent parameters. Furthermore,
the moment of
inertia ${\cal I}$ is kept constant and
chosen as the static moment extrapolated at
zero spin from the experimental data of [17-21]. From
equation (13), we can compute the transition energies
$E_{\gamma}(j) = E(j) - E(j-2)$ and we choose to minimize
$$
\chi^2 = { 1 \over N } \; \sum_j \;
         \left[ { {E_{\gamma}^{\rm th} (j) -
                   E_{\gamma}^{\rm ex} (j)} \over
           {\Delta E_{\gamma}          (j)} } \right]^2
\eqno (17)
$$
where $\Delta E_{\gamma} (j)$ are the experimental errors and $N$
is the number of experimental points entering the fitting procedure.
For the purpose of comparison, the same SD bands have also been
analysed in the framework of the $U_q({\rm su}_2)$ model by
using a similar fitting procedure (same $\chi^2$~;
same moment of inertia~;
the parameter $\gamma$ is kept to the value
$90$\degrees and only the parameter
$\beta'$, with $q = {\rm e}^{ {\rm i} \beta'}$, is freely
varied).

Figure 1 displays the results obtained from the
 $U_{qp}({\rm u }_2)$ and
 $U_{q }({\rm su}_2)$ models. Table 1 shows the
values of the parameters $\beta$ and $\gamma$ for the
    $U_{qp}({\rm u }_2)$ model and of the parameter $\beta'$ for
the $U_{q }({\rm su}_2)$ model.
(Note that the table (Table 1) and
the figure (Figure 1) of this work do not appear in
this version. They can be obtained from the authors.)
For $\gamma = 90$\degrees, the two models coincide and our
results reflect this fact since for $\gamma$ close to
$90$\degrees, as for $^{198}$Pb, we have
$\beta'$ close to
$\beta $ and  the two models give similar
results. However, when $\gamma$ increases, a better agreement
with experiment
is globally obtained for the $U_{qp}({\rm u }_2)$ model~; the
largest discrepancy between the two models is reached in the
case of $^{192}$Hg for which $\gamma= 127$\degrees and
the values of $\beta $ and
              $\beta'$ are quite different.
Finally, it is to be noted that the better agreement for the
$U_{qp}({\rm u }_2)$ model is also depicted by the
values obtained for $\chi$ which range  from 1 to 7    for the
$U_{qp}({\rm u }_2)$ model and from 2 to 1000 for the
$U_{ q}({\rm su}_2)$ model. (In both cases, the large values of
$\chi$ are due to the small values of the experimental errors,
which are of the order of $0.1 \%$ of the transition energies.)

As a further test of our model, we have used the parameters
$\beta$ and $\gamma$ obtained for $^{192}$Hg in order to predict
the transition energies $E_{\gamma}(44)$
                    and $E_{\gamma}(46)$. The so calculated
energies are in good agreement with the new experimental data
obtained with the EUROGAM detector [22].

A few words should be said about the physical interpretation of
the parameters of the $U_{qp}({\rm u }_2)$ model. Like the
parameter $\beta'$ in the
$U_{ q}({\rm su}_2)$ model, the parameter $\beta \sin \gamma$
corresponds to the softness parameter of the VMI model. The
parameter $\beta \cos \gamma$ strongly differentiates
the two models at high spins: a much better
agreement between theory and experiment is generally obtained at high spins
with the $U_{qp}({\rm u }_2)$ model~; this is
especially true for $^{192}$Hg and $^{194}$Hg(1b) for which
the values of $\gamma$ are far from $90$\degrees.
This suggests
that the parameter $\beta \cos \gamma$ manifests itself
at high spins by a weakening in the
increasing (versus spin) of the moment
of inertia.

In conclusion, we have presented an
$U_{qp}({\rm  u}_2)$ model
for rotational spectra of nuclei which presents some advantages
over the
$U_{q }({\rm su}_2)$ model as far as energy levels are
concerned. Our test has been concerned with seven SD bands.
On the other hand, we have also tested our model on the ground
state band of $^{238}$U~; the results  obtained  for this
case study are in good accordance with classical results [23].
Further applications of our model could be
the analysis, for classification purposes,
of identical SD bands: keeping
their moment of inertia as constant, one could extract their
dependency on the $(q,p)$ parameters. Furthermore, it would
be interesting to investigate
$B(E2)$ transition probabilities
in the $qp$-rotor model with
$U_{qp}({\rm u}_2)$ symmetry.
These matters shall be the object of
the thesis by the junior author (R~B).

\sa

\noindent
The authors thank Mich\`ele Meyer and Yuri~F~Smirnov
for very useful comments.
Interesting discussions with Nadine Redon and
members of the $A \sim 190$ EUROGAM
collaboration are also acknowledged. Thanks are due to
Fran\c cois Gieres for a critical reading of the manuscript.

\vfill\eject

\noindent {\bf References}

\sa
\sa

\baselineskip = 0.75 true cm

\noindent \item{[1]}
Raychev P~P, Roussev R~P and Smirnov Yu~F 1990 {\it
J. Phys. G: Nucl. Part. Phys.} {\bf 16} L137

\noindent \item{[2]}
Iwao S 1990 {\it Prog. Theor. Phys.} {\bf 83} 363

\noindent \item{[3]}
Bonatsos D, Argyres E~N, Drenska S~B, Raychev P~P,
Roussev R~P and Smirnov Yu~F 1990 {\it Phys. Lett. B} {\bf 251} 477

\noindent \item{[4]}
Bonatsos D, Drenska S~B, Raychev P~P, Roussev R~P and
Smirnov Yu~F 1991 {\it J. Phys. G: Nucl. Part. Phys.} {\bf 17} L67

\noindent \item{[5]}
Zhilinski\v \i ~B~I and Smirnov Yu~F 1991
{\it Sov. J. Nucl. Phys.} {\bf 54} 10 [Yad. Fiz. {\bf 54} 17]

\noindent \item{[6]}
Bonatsos D, Faessler A, Raychev P~P, Roussev R~P
and Smirnov Yu~F 1992 {\it J. Phys. A: Math. Gen.} {\bf 25} 3275

\noindent \item{[7]}
Bonatsos D, Faessler A, Raychev P~P, Roussev R~P and
Smirnov Yu~F 1992 {\it J. Phys. A: Math. Gen.} {\bf 25} L267

\noindent \item{[8]}
Del Sol Mesa A, Loyola G, Moshinsky M and Vel\'azquez
1993 {\it J. Phys. A: Math. Gen.} {\bf 26} 1147

\noindent \item{[9]}
Sudbery A 1990 {\it J. Phys. A: Math. Gen.} {\bf 23} L697

\noindent \item{[10]}
Chakrabarti R and Jagannathan R 1991 {\it J. Phys. A: Math. Gen.}
{\bf 24} L711

\noindent \item{[11]}
Fairlie D~B and Zachos C~K 1991 {\it Phys. Lett. B} {\bf 256} 43

\noindent \item{[12]}
Schirrmacher A, Wess J and Zumino B 1991 {\it Z. Phys. C}
{\bf 49} 317

\noindent \item{[13]}
Smirnov Yu F and Wehrhahn R F 1992 {\it J. Phys. A:
Math. Gen.} {\bf 25} 5563

\noindent \item{[14]}
Quesne C 1993 {\it Phys. Lett. A} {\bf 174} 19

\noindent \item{[15]}
Meljanac S and Milekovic M 1993 {\it J. Phys. A: Math. Gen.}
{\bf xx} xxxx

\noindent \item{[16]}
Kibler M 1993 {\it Symmetry and Structural
Properties of Condensed Matter} ed W Florek, D Lipinski
and T Lulek (Singapore: World Scientific) p~445

\noindent \item{[17]}
Ye D {\it et al} 1990 {\it Phys. Rev. C} {\bf 41} R13

\noindent \item{[18]}
Riley M A {\it et al} 1990 {\it Nucl. Phys.} {\bf A 512} 178

\noindent \item{[19]}
Henry E A {\it et al} 1991 {\it Z. Phys. A} {\bf 338} 469

\noindent \item{[20]}
Theine K {\it et al} 1990 {\it Z. Phys. A} {\bf 336} 113

\noindent \item{[21]}
Wang T F  {\it et al} 1991 {\it Phys. Rev. C} {\bf 43} R2465

\noindent \item{[22]}
Gall B {\it et al} 1994 Experimental and theoretical
study of the ${\cal I}^{(2)}$ dynamical moment of inertia of
superdeformed bands in the $A=190$ mass region {\it Proc. Int.
Conference on the Future of Nuclear Spectroscopy (Aghia
Pelagia, 1993)} eds W~Gelletly and C~A~Kalfas
(Singapore: World Scientific) to be published

\noindent \item{[23]}
Bohr A and Mottelson B~R 1975 {\it Nuclear Structure,
vol II} (London: Benjamin) p~67

\bye